\documentclass[prl,preprint]{revtex4}
\usepackage{amsmath}
\usepackage{amsfonts}
\usepackage{graphicx}

\begin{document}

\title{
Non-invertibility of Multiple-Scattered QELSS Spectra
 }

\author{George D. J. Phillies}
\email{phillies@wpi.edu}

\affiliation{Department of Physics, Worcester Polytechnic
Institute,Worcester, MA 01609}

\begin{abstract}

We consider the spectrum $S(q,t)$ and field correlation function $f(q,t)$ of light quasielastically scattered from diffusing optical probes in complex viscoelastic fluids.  Relationships between the single-scattering  $f_{1}(q,t)$ and the multiple-scattering $f_{m}(t)$ are examined.  We show that it is fundamentally impossibly to invert $f_{m}(t)$ to obtain $f_{1}(q,t)$ or particle displacement moments $\overline{X^{2n}}$, except with assumptions that are certainly not correct in complex, viscoelastic fluids.  For diffusing dilute probes in viscoelastic fluids, $f_{1}(q,t)$ is determined by all even moments $\overline{X^{2n}}$, $n \geq 1$, of the particle displacement $X$; this information is lost in $f_{m}(t)$.  In the special case of monodisperse probes in a true simple fluid, $f_{1}(q,t)$ can be obtained from $f_{m}(t)$,  but only because the functional form of $f_{1}(q,t)$ is already known.
 
\end{abstract}

\maketitle

For three decades\cite{turner1976aDp}, optical probe diffusion (OPD) has been used to infer the dynamic properties of complex fluids.  In an typical OPD experiment, dilute, intensely-scattering Brownian particles are dispersed in a complex fluid, and quasi-elastic light scattering spectroscopy (QELSS) is used to obtain the spectrum $S(q,t)$ and field correlation function $f_{1}(q,t)$ of the single-scattered light.  Other experimental methods, e.g., fluorescence recovery after photobleaching, have been used to track OPD probes on longer time and distance scales. Probe motions are used to infer dynamic viscosities and relaxation spectra of the complex fluid matrix around the probes. OPD methods have been successfully applied to many complex fluids, including solutions of hydroxypropylcellulose\cite{russo1988bDp}, polyethylene oxide\cite{ullmann1985aDp}, and polyisobutylene\cite{pu1989aDp}, to polymer melts\cite{lin1986aDp}, and to nondilute surfactant solutions\cite{phillies1993a}. 

Recently, QELSS methods have been applied to complex fluids containing concentrated optical probes\cite{maret1987aDp,weitz1993aDp}, in many cases complex fluids also studied with OPD.  These experiments examine light that was profoundly multiply scattered before reaching the detector. Because the same photomultiplier tube/correlator technology is used, OPD and multiple-scattering studies access the same absolute time scales.  However, multiple-scattering spectra may be sensitive to particle motions over shorter distance scales than OPD measurements are. The multiple scattering spectrum arises as a coherent sum over light travelling every allowed path from the laser to the detector.  The allowed paths differ in their lengths, the number of scattering events, and the scattering vector at each scattering event along each path. Individual scattering events along each multiple-scattering path are identical with single-scattering events studied with OPD using QELSS. 

For a single scattering event involving physically dilute probes, $f_{1}(q,t)$ is related to particle positions ${\bf r}_{i}(t)$ by
\begin{equation}
    f_{1}(q,t) = \left\langle \sum_{i=1}^{N} \exp( i {\bf q} \cdot ({\bf r}_{i}(t)- {\bf r}_{i}(0) )  ) \right\rangle
      \label{eq:f1start}
\end{equation}
${\bf q}$ being the scattering vector, the sum being over all $N$ probes, and the dilute-probe condition eliminating terms involving cross-correlations in the positions of pairs of probes. It is convenient to write $\Delta X_{i}(t) = \hat{q} \cdot ({\bf r}_{i}(t)- {\bf r}_{i}(0) )$ for the particle displacement between $0$ and $t$ along the $\hat{q}$ axis. Resummation via cumulants\cite{phillies2005aDp} replaces the overall average with a set of averages over the mean-$2 n^{\rm th}$ particle displacements $\overline{X^{2n}}$, restating $f_{1}(q,t)$ as

\begin{equation}
   f_{1}(q,t) =
    \exp\left[- \left(q^2 \frac{\overline{X^{2}}}{2} - q^{4}\frac{
    (\overline{X^{4}} - 3
    \overline{X^{2}}^{2}) 
    }{24}+     q^{6} \frac{( 30 \overline{X^{2}}^{3} - 15   \overline{X^{2}} \  \overline{X^{4}} + \overline{X^{6}} )}{720} \right) + \ldots \right]
      \label{eq:g1Pexpanded}
\end{equation}
and terms of order $\overline{X^{8}}$ and larger.

In the very special case that the particles are monodisperse, the environments are all the same, and the solvent is a simple fluid with no long-term viscoelastic correlations in the forces on the probes, eq \ref{eq:g1Pexpanded} simplifies to 
\begin{equation}
   f_{1}(q,t) =
    \exp\left[- \left( \frac{1}{2} q^2 \ \overline{X^{2}} \right)\right].
   \label{eq:g1Pexpanded2}
\end{equation}
because in this special case the distribution $P(X)$ of particle displacements is a Gaussian with width $\overline{X^{2}} = 2 Dt$, $D$ being a time-independent diffusion coefficient. 

Fortunately, a simple diagnostic permits the experimenter to determine 
if eq \ref{eq:g1Pexpanded2} is applicable.  Doob\cite{doob1942a} has shown that if the distribution of sequential particle displacements has a joint Gaussian form, so that eq  \ref{eq:g1Pexpanded2} is correct, then the mathematical form of $f_{1}(q,t)$ is fixed: As a purely mathematical consequence $f_{1}(q,t)$ must  precisely a pure exponential $\exp(-\Gamma t)$ in $t$, with $\Gamma$ a time-independent decay rate.  Furthermore, because $P(X)$ is a Gaussian in $X$, $\Gamma$ is necessarily proportional to $q^{2}$.  Contrariwise, if $f_{1}(q,t)$ is not a simple exponential in $t$, or $\Gamma$ does not scale in $q$ as $q^{2}$, then eq \ref{eq:g1Pexpanded} does not simplify to \ref{eq:g1Pexpanded2}, and from Doob's theorem $P(X)$ is not Gaussian in $X$. 

What is the actual nature of probe motion in polymer solutions?  Streletzky\cite{somestreletzky} reports extensive 
light scattering spectra for mesoscopic polystyrene sphere probes diffusing in hydroxypropylcellulose(HPC):water.  Manifestly, $f_{1}(q,t)$ for probes in HPC:water is not a pure exponential.  Instead, $f_{1}(q,t)$ is a sum of two stretched exponentials in $t$.   From Doob's theorem, eq \ref{eq:g1Pexpanded2}, and results derived by assuming it to be correct, do not apply in this typical probe:polymer system.

Figure 1 shows how the mean relaxation rates\cite{otherstreletzky} of the two QELSS modes of 20nm probes in HPC:water depend on $q$. One mode is simply diffusive, with a relaxation rate that scales as $q^{2}$.  For the other mode, the mean relaxation rate has a complicated $q$-dependence, not the $q^{2}$-dependence of eq \ref{eq:g1Pexpanded2}.  In this typical probe:polymer system, $\log(f_{1}(q,t))$ is also not linear in $q^{2}$.

The actual distribution of particle displacements for probes in water:glycerol and protein solutions was measured by Apgar, Tseng, and co-workers\cite{apgar2000a} using video microscopy.  In water:glycerol, $P(X)$ is Gaussian in $X$. In concentrated actin and actin:fascin solutions, $P(X)$ is profoundly non-Gaussian. 

By comparison with these experiments\cite{somestreletzky,otherstreletzky,apgar2000a}, Doob's theorem shows that probe diffusion in representative polymer solutions is described by eq \ref{eq:g1Pexpanded}, and not by eq \ref{eq:g1Pexpanded2}.  

For the multiple-scattering spectrum, the field correlation function $f_{m}(t)$ has been proposed to be given by\cite{weitz1993aDp}
\begin{equation}
      f_{m}(t) = \int ds P(s) \exp(-2 k_{o}^{2} \overline{X(t)^{2}} s/\ell^{*})
      \label{eq:mss1}
\end{equation}
Here $ds$ averages over the distribution $P(s)$ of pathlengths $s$, and $k_{o}$ and $\ell^{*}$ are the illuminating light wavevector and photon effective mean free path, respectively. In this equation, the term $2 k_{o}^{2}$ arises as the angular average over allowed scattering vectors $q^{2}$, based on the assumption that $\log(f_{1}(q,t) \sim \exp(- D q^{2} t)$.  Durian\cite{durian1995a} has shown a path to improving approximations in eq \ref{eq:mss1}.  This paper is not concerned with the accuracy of those approximations.

With respect to this paper, the only important issue in eq \ref{eq:mss1} is that $f_{m}(t)$ is obtained as an intricate average of $f_{1}(q,t)$ over allowed scattering vectors $q$ and other variables. $f_{m}(t)$ therefore has no residual $q$-dependence.  (The dependence of $f_{m}(t)$ on window face arises from the dependence of $P(s)$ on the detector location, not from the $q$-dependence of $f_{1}(q,t)$.)  In contrast,  $f_{1}(q,t)$ is a nontrivial function of two variables, $t$ and $q$.

The objective in analyzing a multiple-scattering QELSS spectrum is to infer from $f_{m}(t)$ the single-scattering field correlation function $f_{1}(q,t)$.  However, $f_{m}(t)$ is a function of a single continuous variable $t$, while $f_{1}(q,t)$ is a nontrivial function of two continuous variables $t$ and $q$.  It is fundamentally impossible to invert a function of one variable, obtained as an average over an underlying two-variable function, to recover the two-variable function.  Therefore, the goal of using a spectrum of multiple-scattered light to infer the spectrum of single-scattered light is {\em in the general case} unattainable.

In one special case, $f_{1}(q,t)$ can be recovered from $f_{m}(t)$, namely the case in which the $q$-dependence of $f_{1}(q,t)$ is known {\em a priori}.  For example, for probe particles in simple solvents such as water:glycerol, $f_{1}(q,t)$ is known from QELSS determinations to be a pure exponential in $q^{2}$.  This {\em a priori} knowledge of $f_{1}(q,t)$ effectively reduces $f_{1}(q,t)$ to a one variable function.  There is no fundamental obstacle--the details may be tedious--to recovering a one-variable function from a different function of the same variable. Of course, if $f_{1}(q,t)$ is already known by direct means, indirect paths to its determination appear redundant. Unfortunately, 'probe particles in simple solvents' are the systems likely to be used to test methods for recovering  $f_{1}(q,t)$ from $f_{m}(t)$.  Such tests do not duplicate key features of probe motion in viscoelastic fluids, becuse probes in simple solvents do not have the non-Gaussian displacement distribution and corresponding non-trivial $q$-dependence of $f_{1}(q,t)$ found for probes in complex viscoelastic fluids.

It might in principle be possible to construct nanostructured probes having the property that their static scattering factor $S(q)$ is strongly peaked over a narrow range of angles.  The multiple-scattering spectrum of these hypothetical probes would sample $f_{1}(q,t)$ over a narrow range in $q$.  Use of a series of probes with different preferred scattering angles in a series of experiments on the same complex fluid might then allow reconstruction of the complete $f_{1}(q,t)$ from $f_{m}(t)$.

For probes in viscoelastic complex fluids the functional form of $f_{1}(q,t)$ is not known {\em a priori}.  The functional dependences of $f_{1}(q,t)$ on $t$ and $q$ then both need to be determined experimentally.  Multiple scattering spectra have averaged out the $q$-dependence, and therefore cannot be used to  reconstruct $f_{1}(q,t)$.  In contrast, QELSS/OPD single-scattering experiments have long\cite{turner1976aDp} routinely determined $f_{1}(q,t)$ directly and unambiguously.

\pagebreak

\centerline{\bf Figure Captions}

Figure 1. $q^{2}$-dependence of the inverse mean
relaxation times $\Gamma$ of the modes of $f_{1}(q,t)$ for 20nm probes in 4g/l hydroxypropylcellulose,
after ref.\ \cite{otherstreletzky}

\pagebreak

\pagebreak

\centerline{\Large Phillies, 1}

\vfill

\begin{figure} 

\includegraphics{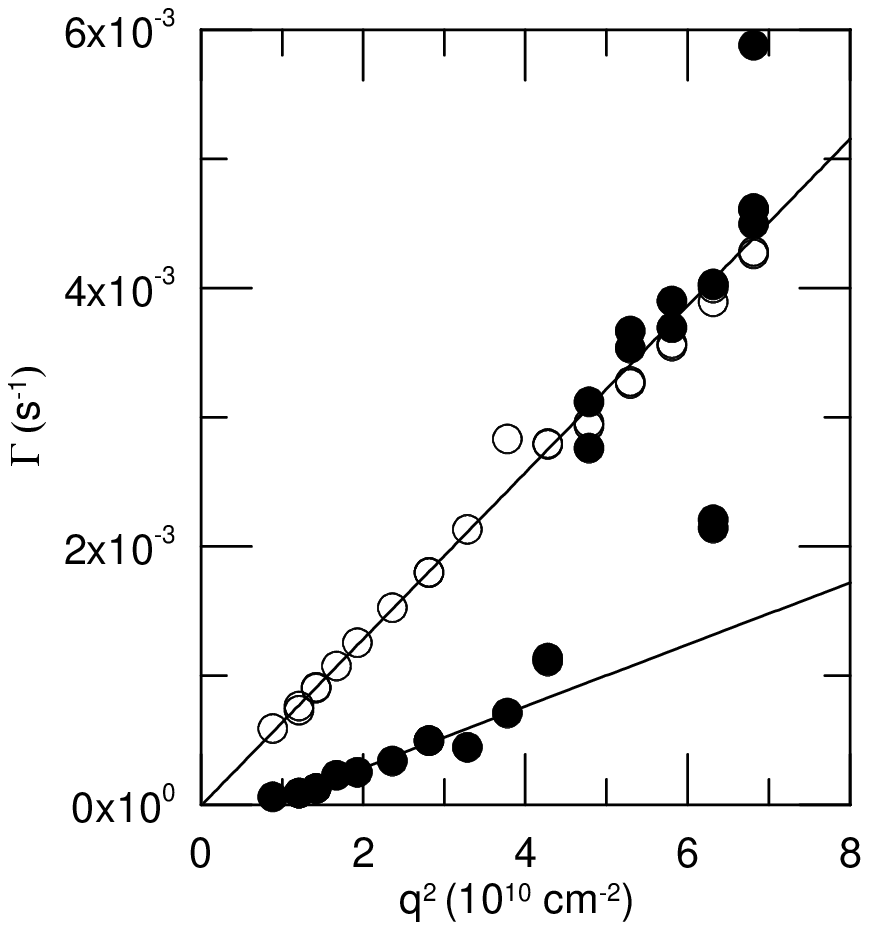} 

\end{figure}

\end{document}